\documentclass{biometrika}

\usepackage{times}
\usepackage{bm}
\usepackage{amsmath, amssymb}
\usepackage{natbib}
\usepackage{pst-plot, pst-node, pst-text}
\usepackage{graphicx, subfigure}

\RequirePackage[colorlinks,citecolor=blue,urlcolor=blue]{hyperref}
\RequirePackage{hypernat}

\def\P{\rm pr}
\newenvironment{proof_of}[1][]{\begin{trivlist}
\item[\hskip \labelsep {\bfseries Proof of #1.}]}{\end{trivlist}}

\def\convprob{\rightarrow_p}

\def\R{\mathbb R}

\def\point{\hspace{-.08cm}\cdot\hspace{-.08cm}}
\newcommand{\ind}   [1]{1( #1 ) }
\newcommand{\st}[0]{:}

\begin{document}

\jname{Biometrika}


\markboth{M. H. Maathuis \and M. G. Hudgens}{Competing risks current status data}

\title{Nonparametric inference for competing risks current status data with continuous, discrete or grouped observation times}

\author{M. H. Maathuis}
\affil{Seminar f\"ur Statistik, ETH Z\"urich, R\"amistrasse 101, 8092 Z\"urich, Switzerland \email{maathuis@stat.math.ethz.ch}}

\author{M. G. Hudgens}
\affil{Department of Biostatistics, University of North Carolina at Chapel Hill, 3107-E McGavran-Greenberg Hall, Chapel Hill, North Carolina 27599, U.S.A.
 \email{mhudgens@bios.unc.edu}}

\maketitle

\begin{abstract}
    New methods and theory have recently been developed to nonparametrically estimate cumulative
    incidence functions for competing risks survival data subject to current status censoring. In
    particular, the limiting distribution of the nonparametric maximum likelihood estimator and a simplified naive estimator have
    been established under certain smoothness conditions. In this paper, we establish the large-sample
    behavior of these estimators in two additional models, namely when the observation time distribution has discrete support and when the observation times are grouped.
    These asymptotic results are applied to the construction of confidence intervals in the three different models. The methods are illustrated on two data sets regarding the cumulative incidence of different
    types of menopause from a cross-sectional sample of women in the United States and of
    subtype-specific HIV infection from a sero-prevalence study in injecting drug users in Thailand.
\end{abstract}


\begin{keywords}
   Competing risk; Confidence interval; Current status data; HIV-prevalence; Interval censoring; Limiting distribution; Nonparametric maximum likelihood estimator
\end{keywords}

\section{Introduction}\label{sec: introduction}

Current status data with competing risks arise in cross-sectional studies that assess the current status of
individuals in the sample with respect to an event that can be caused by several mechanisms. An example is Cycle I of the Health Examination Survey in the United States \citep{MacMahonWorcestor66}. This study recorded the age and menopausal status of the female participants, where menopausal status could be pre-menopausal, post-menopausal
due to an operation, or post-menopausal due to natural causes. Based on these data, the cumulative incidence of natural and operative menopause can be estimated as a function of age. A second example is the Bangkok Metropolitan Administration injecting drug users
cohort study \citep{KitayapornEtAl98, VanechseniEtAl01}. This study recorded the age and HIV status of
injecting drug users, where HIV status could be HIV negative, HIV positive with subtype B, HIV positive with subtype E, or HIV positive with some other subtype. Based on these data, the subtype-specific cumulative incidence of HIV can be estimated as a function of age.

New methods and theory
have recently been developed to nonparametrically estimate cumulative incidence functions based on current status data with competing risks.
\citet{HudgensSattenLongini01} and \citet{JewellVanderLaanHenneman03} derived and studied the nonparametric maximum likelihood and also introduced several other estimators, including the so-called naive estimator of \citet{JewellVanderLaanHenneman03}. \cite{Maathuis06thesis} and
\citet{GroeneboomMaathuisWellner08a, GroeneboomMaathuisWellner08b} derived the large-sample behavior of the maximum likelihood estimator and the naive estimator in a smooth model that imposes certain smoothness conditions on the cumulative incidence functions and the observation time distribution. In this model, the local rate of convergence of the maximum likelihood estimator is $n^{1/3}$ \citep[Theorem $4 \point 17
$]{GroeneboomMaathuisWellner08a}, slower than the usual $n^{1/2}$ rate. Moreover, its limiting distribution is non-standard and involves a self-induced system of slopes of convex minorants of Brownian motion processes plus parabolic drifts \citep[Theorems $1\point 7$ and $1 \point 8$]{GroeneboomMaathuisWellner08b}. The naive estimator has the same local rate of convergence as the maximum likelihood estimator, but its limiting distribution is simpler, since it does not involve a self induced system \citep[Theorem $1 \point 6$]{GroeneboomMaathuisWellner08b}.

In practice, recorded observation times are often discrete, making the smooth model unsuitable. We therefore study the large sample behavior of the maximum likelihood estimator and the naive estimator in two additional models: a discrete model  in which the observation time distribution has discrete support, and a grouped model in which the observation times are assumed to be rounded in the recording process, yielding grouped observation times.

We show that the large sample behavior of the estimators in the discrete model is fundamentally different from that in the smooth model: the maximum likelihood estimator and the naive estimator converge locally at rate $n^{1/2}$, and their limiting distributions are identical and normal. These results are related to the work of \citet{YuSchickLiWong98-case1}, who studied the asymptotic behavior of the maximum likelihood estimator for current status data with discrete observation times in the absence of competing risks. There are also connections to unpublished work of Tang, Banerjee and Kosorok, who studied the limiting distribution of the maximum likelihood estimator for current status data when the observation times fall on a grid that depends on the sample size.

The grouped model is related to the work of \cite{WoodroofeZhang99} and \cite{ZhangKimWoodroofe01}, who
considered the maximum likelihood estimator for a nondecreasing density when the observations are grouped. We are not aware,
however, of any work on the maximum likelihood estimator for interval censored data with grouped observation times, even though such grouping frequently occurs in
practice. For example, in the menopause data the ages of the women were grouped in the intervals $(25,30]$, $(30,35]$, $(35,36]$, $(36,37]$, $\dots$, $(58,59]$ and recorded as the midpoints of these intervals. The menopausal status, on the other hand, was determined at the
exact but unrecorded time of interview, yielding a mismatch between the recorded status and the recorded observation time. For example, if a 30.7 year old pre-menopausal woman is interviewed, she is recorded as pre-menopausal with rounded age 32.5. When ignoring the rounding, as done in previous analyses of these data, this is taken to mean that she was interviewed at age 32.5 and that she was pre-menopausal at that age. A correct interpretation of the data is, however, that she was pre-menopausal at some unknown age in the interval $(30,35]$. In particular, the data do not reveal her menopausal status at age 32.5; in actuality, she might have been post-menopausal at that age, for example due to an operation.

The grouped model accounts for such grouping of observation times. We show that the likelihood in this model can be written in the same form as in the discrete model, but in terms of different parameters, representing weighted averages of the cumulative
incidence functions over the grouping intervals, where the weights are determined by the observation time distribution. This similarity with the  discrete model implies that the maximum likelihood estimator and the naive estimator in the grouped model can be computed with existing software, and that their limiting distributions can be derived as in the discrete model. However, since the likelihood is written in terms of different parameters, the estimates under the grouped model must be interpreted differently.
The ideas incorporated in the grouped model can be easily extended to other forms of interval censored data.

The asymptotic results in the three models are applied to the construction of confidence intervals, a problem that has received little attention until now. In the discrete and grouped models, confidence intervals can be constructed by standard methods, for example using the bootstrap or the limiting distributions derived in this paper.
In the smooth model, the non-standard limiting behavior of the estimators makes the construction of confidence intervals less straightforward. In this case, we advocate using likelihood ratio confidence intervals \citep{BanerjeeWellner01} based on the naive estimator.

\section{Models}\label{sec: model}

\subsection{Exact observation times}\label{sec: model exact observation times}

Consider the usual competing risks
setting where an event can be caused by $K$ competing risks, with $K\in \{1,2,\dots\}$ fixed. The random
variables of interest are $(X,Y)$, where $X \in \R$ is the time of the event of interest, and
$Y\in\{1,\dots,K\}$ is the corresponding cause. The goal is to estimate the cumulative incidence
functions $F_0 = (F_{01},\dots,F_{0K})$, where $F_{0k}(t) =
\P(X\le t, Y=k)$ for $k=1,\dots,K$. The cumulative incidence functions are non-negative,
monotone non-decreasing, and satisfy $\sum_{k=1}^K F_{0k}(t) = \P(X\le t) \le 1$.

The difficulty in estimating the cumulative incidence functions is that we cannot observe $(X,Y)$
directly. Rather, we observe the current status of a subject at a single random observation time
$C \in \R$. Thus, at time $C$ we observe whether or not the event of interest has
occurred, and if and only if the event has occurred, we also observe the cause $Y$. We assume $C$ is independent of  $(X,Y)$. Let $G$ denote the
distribution of $C$, and let $(C, \Delta)$ denote the observed data, where
$\Delta=(\Delta_1,\dots,\Delta_{K+1})$ is an indicator vector for the status of the subject at time $C$,
\begin{align}\label{eq: def Delta}
   \vspace{-.1cm}
   \begin{array}{rl}
   \Delta_k & = \ind{X\le C, Y=k}, \,\,\,k=1,\dots,K,\\
   \Delta_{K+1} & = \ind{X>C},
   \end{array}
\end{align}
where $\ind(\cdot)$ is the indicator function. 
To make this concrete, consider the HIV data discussed in Section \ref{sec: introduction}, where $X$
is the age at HIV infection, $C$ is the age at screening, and there are $K=3$
competing risks representing the HIV subtypes: $Y=1$ for subtype B, $Y=2$ for subtype E, and $Y=3$
for other subtypes.

We consider the maximum likelihood estimator for $F_0$ based on $n$ independent and identically distributed observations
of $(C,\Delta)$, denoted by $(C_i,\Delta^i)$, $i=1,\dots,n$, where
$\Delta^i = (\Delta_1^i,\dots,\Delta_{K+1}^i)$. For any $K$-tuple $(x_1,\dots,x_K)$ let $x_+ = \sum_{k=1}^K x_k$ and, unless otherwise defined, let $x_{K+1} = 1-x_+$. Moreover, define the set
$\mathcal F_K = \{F=(F_1, . . . , F_K) : F_1,\dots,F_K$ are cumulative incidence functions and $F_+(t)\le 1$ for all $t \in \R \}$. A maximum likelihood estimator for $F_0$ is defined as any $\hat F_n =(\hat F_{n1},\dots,\hat F_{nK}) \in \mathcal F_K$
satisfying $l_n(\hat F_n) =
\max_{F \in \mathcal F_K} l_n(F)$, where $l_n(F)$ is the log likelihood
\begin{align}\label{eq: def ln(F)}
  \vspace{-.1cm}
  l_n(F) = \frac{1}{n}\sum_{i=1}^n \sum_{k=1}^{K+1} \Delta_k^i \log \{F_k(C_i)\},
\end{align}
with the convention $0\log 0=0$; see also \citet{JewellVanderLaanHenneman03}, equation (1).

We also consider the naive estimator $\tilde F_n = (\tilde F_{n1},\dots,\tilde F_{nK})$ of  \citet{JewellVanderLaanHenneman03}, whose $k$th component is defined as any $\tilde F_{nk} \in \mathcal F_1$ satisfying $l_{nk}(\tilde F_{nk}) = \max_{F_k \in \mathcal F_1} l_{nk}(F_k)$,
where
\begin{align}\label{eq: def lnk(Fk) naive}
   \vspace{-.1cm}
   l_{nk}(F_k) = \frac{1}{n}\sum_{i=1}^n \left[ \Delta_k^i \log \{F_k(C_i)\} + (1-\Delta_k^i) \log \{1-F_k(C_i)\} \right]
\end{align}
is the marginal log likelihood for the reduced current status data $(C_i,\Delta_k^i)$, $i=1,\dots,n$, and $\mathcal F_1$ is obtained from $\mathcal F_K$ by taking $K=1$.
Since $\tilde F_{nk}$ only uses the $k$th entry of the $\Delta$-vector, the naive estimator
splits the estimation problem into $K$ well-known univariate current
status problems. Therefore, its computation and asymptotic theory follow
straightforwardly from known results on current status data. But this simplification comes
at a cost. For example, $\tilde F_{n+}$ need not be bounded by one, and the naive estimator
has been empirically shown  to be less efficient than the maximum likelihood estimator in the smooth model
\citep{GroeneboomMaathuisWellner08b}.

The R-package {\tt MLEcens} provides an efficient and stable method to compute the maximum likelihood estimator. This algorithm first uses the Height Map Algorithm of \cite{Maathuis05} to compute the areas to which the maximum likelihood estimator can possibly assign probability mass, called maximal intersections. Next, it computes the
amounts of mass that must be assigned to the maximal intersections. This involves solving a high-dimensional convex optimization problem, which is done using the support reduction algorithm of \citet{GroeneboomJongbloedWellner08}.
\citet{JewellKalbfleisch04} describe an alternative algorithm for the computation of the MLE, based on the pool adjacent violators algorithm of \citet{AyerEtAl55}.

The maximum likelihood estimator and the naive estimator are not defined uniquely at all times. \citet{GentlemanVandal02} defined two types of non-uniqueness for estimators based on censored data: mixture non-uniqueness and representational non-uniqueness. Mixture non-uniqueness occurs when the probability masses assigned to the maximal intersections are non-unique. Representational non-uniqueness refers to the fact that the estimator is indifferent to the distribution of mass within the maximal intersections. The maximum likelihood estimator for current status data with competing risks is always mixture unique \citep[Theorem $2\point 20$]{Maathuis06thesis}, and mixture uniqueness of the naive estimator follows as a special case of this. One can account for representational non-uniqueness of the estimators by providing a lower bound that assigns all mass to the right endpoints of the maximal intersections, and an upper bound that assigns all mass to the left endpoints of the maximal intersections.

\subsection{Exact observation times with discrete support}

Section \ref{sec: model exact observation times} does not impose any assumptions on the observation time distribution $G$, and hence is valid for both continuous and discrete observation times. However, the formulas can be simplified when $G$ is discrete. In this case, let $G(\{s\})$ denote the point mass of $G$ at $s$, and let $\mathcal S = \{s \in \R\st G(\{s\})>0\}$ denote the support of $G$, where $\mathcal S$ is countable but possibly infinite. Defining
\begin{align*}
   N_k(s) & = \frac{1}{n}\sum_{i=1}^n \Delta_k^i \ind{C_i=s}, \qquad k=1,\dots,K+1, \, s\in \mathcal S,
\end{align*}
and $N(s) = \sum_{k=1}^{K+1} N_k(s)$, the log likelihood \eqref{eq: def ln(F)} reduces to
\begin{align}\label{eq: llh Nkj}
   l_n(F) = \sum_{s\in \mathcal S} \sum_{k=1}^{K+1} N_{k}(s)\log \{ F_k(s)\},
\end{align}
and the marginal log likelihood \eqref{eq: def lnk(Fk) naive} for the naive estimator
becomes
\begin{align*}
   l_{nk}(F_k) = \sum_{s \in \mathcal S} \left[ N_{k}(s) \log \{ F_k(s)\} + \{N(s) - N_{k}(s)\}\log\{1-F_k(s)\}\right].
\end{align*}
The spaces $\mathcal F_K$ and $\mathcal F_1$ can also be simplified, as the nonnegativity, monotonicity and boundedness constraints only need to hold at points $s\in \mathcal S$.


\subsection{Grouped observation times}\label{sec: model grouped}

In many applications, only rounded versions of the observation times are recorded, yielding grouped observation times. We introduce a new model for this type of data, called the grouped model. For any interval $I$ on the real line, define $G(I)=\int_{c\in I}dG(c)$. Let $\mathcal I$ be
a countable but possibly infinite set of mutually exclusive intervals such that $G(I)>0$ for all $I\in \mathcal I$. For each  $I\in \mathcal I$, let $m(I)$ denote a unique point in the interval, for example its midpoint, and let $\mathcal M = \{ m(I) \in \R \st I\in \mathcal I\}$. For each $m\in \mathcal M$, let $I(m)$ denote the corresponding interval in $\mathcal I$.

The observation scheme in the grouped model is as follows. As before, the current status of a subject is assessed at a single random time $C \in \R$, where $C$ is independent of $(X,Y)$. The difference
is, however, that we no longer observe $C$. Instead, all observation times falling into interval $I$ are grouped and rounded to $m(I)$.
Thus, the observed data are $(D, \Delta)$, where $D = \sum_{I\in \mathcal I} m(I) \ind{C\in I}$ is the rounded version of $C$, and
$\Delta$ is the indicator vector corresponding to the status of the subject at the exact time $C$, as defined in \eqref{eq: def Delta}. We study the maximum likelihood estimator
and the naive estimator based on $n$ independent and identically distributed observations of $(D,\Delta)$, which we denote by $(D_i,\Delta^i)$,
$i=1,\dots,n$.

To derive the likelihood in the grouped model, we compute $\P(D=d, \Delta=\delta)$ for
$d\in \mathcal M$ and $\delta \in
\{e_1,\dots,e_{K+1}\}$, where $e_k$ is the unit vector in $\R^{K+1}$ with
a $1$ at the $k$th entry. Conditioning on the exact observation time $C$ yields
\begin{align}
   & \P(D =d, \Delta=\delta) \notag \\
    & \qquad = \int \P(D=d, \Delta=\delta \mid C=c)dG(c) = \int_{c\in I(d)} \P(\Delta=\delta \mid C=c)dG(c) \notag \\
    & \qquad = \prod_{k=1}^{K+1} \left\{ \int_{c\in I(d)} F_{0k}(c)dG(c)\right\}^{\delta_k}
                          = G\{I(d)\} \prod_{k=1}^{K+1} \left[ H_{0k}\{I(d)\} \right]^{\delta_k} ,
                          \label{eq: prob grouped}
\end{align}
where
\begin{align*}
   H_{0k}\{I(d)\} = [ G\{I(d)\} ]^{-1} \int_{c\in I(d)} F_{0k}(c)dG(c), \,\,k=1,\dots,K
\end{align*}
and $H_{0,K+1}\{I(d)\} = 1-H_{0+}\{I(d)\}$
are weighted averages of $F_{01}, \dots, F_{0,K+1}$ over $I(d)$ with weights determined by $G$. It is convenient to work with these weighted averages, as they must obey the same constraints as the cumulative incidence functions.
More precisely, considering $H_{0k}$, $k=1,\dots,K$, as functions that maps $m$ to $H_{0k}\{I(m)\}$, the constraints on $F_{01},\dots,F_{0K}$ imply that $H_{01}, \dots,H_{0K}$ must be non-negative and non-decreasing and satisfy $H_{0+}\{I(m)\} \le 1$ for all $m\in \mathcal M$. Let $\mathcal H_K$ denote the space of such allowable $K$-tuples $(H_1,\dots,H_K)$.

The term $G\{I(d)\}$ in the right hand side of \eqref{eq: prob grouped} can be dropped from the likelihood, as it does not depend on $F$.
Hence, a maximum likelihood estimator for $H_0 = (H_{01},\dots,H_{0K})$ is defined as any $\hat H_n \in \mathcal H_K$ satisfying
$l_n^{\rm group}(\hat H_n) = \max_{H\in \mathcal H_K} l_n^{\rm group}(H)$, where
\begin{align}\label{eq: def ln grouped}
   l_{n}^{\rm group}(H) = \frac{1}{n} \sum_{i=1}^n \sum_{k=1}^{K+1} \Delta_k^i \log[ H_k\{I(D_i)\}].
\end{align}
Expression \eqref{eq: def ln grouped} has the same form as \eqref{eq: def ln(F)}, but with $F_k(C_i)$
replaced by the weighted average $ H_k\{I(D_i)\}$. As in the discrete model, \eqref{eq: def ln grouped}
can be simplified further:
\begin{align}\label{eq: llh Mkj}
   l_{n}^{\rm group}(H) = \sum_{I\in \mathcal I} \sum_{k=1}^{K+1} M_{k}(I) \log \{H_k(I)\},
\end{align}
where
\begin{align*}
   M_{k}(I) & = \frac{1}{n}\sum_{i=1}^n \Delta_k^i \ind{D_i=m(I)}, \qquad k=1,\dots,K+1, \,I\in \mathcal I.
\end{align*}


Since the log likelihood \eqref{eq: llh Mkj} has the same form as \eqref{eq: llh Nkj}, and also the constraints on the maximization problems for the discrete and grouped models are equivalent, the maximum likelihood estimator in the grouped model can be computed with existing software. Moreover, its asymptotic theory follows straightforwardly from the theory for the discrete model. The important difference between the two models is, however, that the resulting estimates must be interpreted differently. In the discrete model, one estimates the cumulative incidence functions at points $s\in \mathcal S$. In the grouped model, the cumulative incidence functions are unidentifiable in general, and one estimates the weighted averages of the cumulative incidence functions over intervals $I\in \mathcal I$.

The naive estimator $\tilde H_n$ in the grouped model can be derived analogously. Defining $M(I) = \sum_{k=1}^{K+1} M_{k}(I), I\in \mathcal I$, the marginal log likelihood for the $k$th component is
\begin{align}\label{eq: llh Mkj naive}
   l_{nk}^{\rm group}(H_k) = \sum_{I\in \mathcal I} \left[ M_{k}(I)\log \{H_k(I)\} + \{M(I) - M_{k}(I)\}\log\{1-H_k(I)\}\right],
\end{align}
and $\tilde H_{nk}\in \mathcal H_1$ is defined by $l_{nk}^{\rm group}(\tilde H_{nk}) = \max_{H_k\in \mathcal H_1} l_{nk}^{\rm group}(H_k)$.

\begin{remark}\label{remark: grouped model}
    In general, $F_{0k}(m) \neq H_{0k}(I(m))$, but equality can occur in special situations. For example, $F_{0k}(m)=H_{0k}(I(m))$ if $F_{0k}$ is constant on $I(m)$, if both $F_{0k}$ and $G$ are linear on $I(m)$ and $m$ is the midpoint of $I(m)$, or if the only mass of $G$ on $I(m)$ consists of a point mass at $m$. The latter shows that the grouped model generalizes the discrete model.
\end{remark}

\section{Local asymptotics of the estimators}
\label{sec: asymptotics}

\subsection{Strong consistency in the discrete and grouped models}

The maximum likelihood estimator and the naive estimator are Hellinger consistent when the observation times are recorded exactly, for any observation time distribution $G$ \citep[Theorem $4 \point 6$]{Maathuis06thesis}.
Using the equivalence between Hellinger distance and total variation distance, this implies consistency in total variation \citep[Corollary $4 \point 7$]{Maathuis06thesis}, which in turn implies strong pointwise consistency at all points $s\in \mathcal S$ in the discrete model, as stated in Theorem \ref{th: consistency discrete}.
\begin{theorem}\label{th: consistency discrete}
   \citep[Corollary $4 \point 9$]{Maathuis06thesis} In the discrete model, $\hat F_{nk}(s) \to F_{0k}(s)$ and $\tilde F_{nk}(s)\to F_{0k}(s)$ almost surely as $n\to \infty$ for all $s\in \mathcal S$.
\end{theorem}
Since the form of the log likelihood and the constraints on the allowable functions are identical in the discrete and grouped models, the proofs for the discrete model carry over directly to the grouped model. This leads to Theorem \ref{th: consistency grouped}, which we give without proof.
\begin{theorem}\label{th: consistency grouped}
   In the grouped model, $\hat H_{nk}(I) \to H_{0k}(I)$ and $\tilde H_{nk}(I)\to H_{0k}(I)$ almost surely as $n\to \infty$ for all $I\in \mathcal I$.
\end{theorem}

\subsection{Limiting distributions in the discrete model}\label{sec: large-sample local discrete}

Denote the infimum and supremum of $\mathcal S$ by $s_{\inf}$ and $s_{\sup}$. Define
$s_- = \sup\{x\in \mathcal S \st x<s\}$ for $s\in \mathcal S$ with $s\neq s_{\inf}$, and 
$s_+ = \inf\{x\in \mathcal S \st x>s\}$ for $s\in \mathcal S$ with $s\neq s_{\sup}$.
Define $s \in \mathcal S$ to be a regular point if $F_{0k}(s)=0$ for all $k=1,\dots,K$ or the following two conditions hold: (i) if $s\neq s_{\inf}$ then $s_- \in \mathcal S$ and for each $k=1,\dots,K$ either $F_{0k}(s_-)<F_{0k}(s)$ or $F_{0k}(s)=0$, and (ii) if $s\neq s_{\sup}$ then $s_+ \in \mathcal S$ and for each $k=1,\dots,K$ either $F_{0k}(s) < F_{0k}(s_+)$ or $F_{0k}(s)=0$. If $\mathcal S$ is a finite set and $s\in \mathcal S \setminus \{s_{\inf},s_{\sup}\}$, then $s_-$ and $s_+$ are simply the points directly to the left and right of $s$, and conditions (i) and (ii) are equivalent to requiring that for each $k=1,\dots,K$ either $F_{0k}(s_-)<F_{0k}(s)<F_{0k}(s_+)$ or $F_{0k}(s)=0$. As a second example, suppose that $\mathcal S$ is the set of rational numbers. Then for any point $s\in \mathcal S$ we have $s\notin \{ s_{\inf}, s_{\sup}\}$ and $s_- = s = s_+$. Hence, conditions (i) and (ii) are only satisfied if $F_{0k}(s)=0$ for all $k=1,\dots,K$. \citet{YuSchickLiWong98-case1} introduced regular points in the current status model without competing risks. Our definition generalizes theirs by allowing for competing risks. Moreover, we allow the parameters to be on the boundary of the parameter space. For example, $s\in \mathcal S$ can be a regular point when $F_{0k}(s)=0$ for some or all of the $F_{0k}$'s, and $s=s_{\sup}$ can be a regular point when $\sum_{k=1}^K F_{0k}(s)=1$ or when $F_{0k}(s)=\lim_{t\to\infty} F_{0k}(t)$ for some of the $F_{0k}$'s.

We now introduce the following simple estimator for $F_{0k}(s)$:
   $$\breve F_{nk}(s) = N_{k}(s)/N(s), \qquad k=1,\dots,K, s\in \mathcal S,$$
where we set $0/0=0$. This estimator is very simple, in the sense that $\breve F_{nk}$ does not obey monotonicity constraints and uses only the $k$th component of the $\Delta$-vector. Lemma \ref{lemma: breve Fn} below states that $\breve F_n$ is the maximum likelihood estimator for $F_0$ if the monotonicity constraints on the cumulative incidence functions are discarded. Next, Lemma \ref{th: MLE=multinomial} establishes that for any regular point $s\in \mathcal S$, $\hat F_n(s)=\tilde F_n(s)=\breve F_n(s)$ with probability tending to one as $n \to \infty$. Hence, at such points the limiting distributions of $\hat F_n(s)$ and $\tilde F_n(s)$ equal the limiting distribution of $\breve F_n(s)$. This yields asymptotic normality of $\hat F_n(s)$ and $\tilde F_n(s)$ at regular points, as stated in Theorem \ref{th: normal limit}. All proofs are deferred to Section \ref{sec: proofs}.

\begin{lemma}\label{lemma: breve Fn}
   Let $\mathcal F_K^* = \{F=(F_1, . . . , F_K) : F_k(s) \ge 0$ for $k = 1,\dots,K$ and $F_+(s)\le 1$ for all $s\in \mathcal S\}$.
    Then $l_n(\breve F_n) \ge l_n(F)$ for all $F \in \mathcal F_K^*$, and $l_n(\breve F_n) > l_n(F)$ for all $F\in \mathcal F_K^*$ such that $F(s) \neq \breve F_n(s)$ for some $s\in \mathcal S$ with $N(s)>0$.
\end{lemma}

\begin{lemma}\label{th: MLE=multinomial}
   For any regular point $s\in \mathcal S$ in the discrete model,
   $$\P\{\hat F_{n}(s) = \tilde F_{n}(s) = \breve F_{n}(s)\} \to 1, \,\,\, \quad n\to \infty.$$
\end{lemma}

\begin{theorem}\label{th: normal limit}
   For any regular point $s \in \mathcal S$ in the discrete model,
   \vspace{-.1cm}
  \begin{align*}
      n^{1/2}\{ \hat F_n(s) - F_0(s)\} = n^{1/2} \left( \begin{array}{c}
          \hat F_{n1}(s)-F_{01}(s) \\
          \vdots\\
          \hat F_{nK}(s)-F_{0K}(s)
          \end{array}\right)
   \end{align*}
   is asymptotically normal with mean zero and covariance matrix $V(s)$, where $V(s)$ is a $K\times K$ matrix with entries
   \vspace{-.05cm}
   \begin{align*}
      \{ V(s)\}_{k,\ell} & = \left[ F_{0k}(s)\ind{k=\ell} - F_{0k}(s)F_{0\ell}(s) \right]/G(\{s\}), \quad  k,\ell \in \{1,\dots,K\}.
   \end{align*}
   For any finite collection of regular points $s_1,\dots,s_p$ in $\mathcal S$, the stacked vector $n^{1/2}\{ \hat F_n(s_1)-F_0(s_1), \dots, \hat F_n(s_p)-F_0(s_p)\}$ is asymptotically normal with mean zero and block diagonal covariance matrix with blocks $V(s_1),\dots,V(s_p)$. Consistent estimators for the elements of $V(s)$, $s\in \mathcal S$, are
   \vspace{-.05cm}
   \begin{align*}
      \{\hat V_n(s)\}_{k,\ell} = [\hat F_{nk}(s)\ind{k=\ell} - \hat F_{nk}(s)\hat F_{n\ell}(s) ]/N(s), \quad k,\ell\in \{1,\dots,K\}.
   \end{align*}
   The same results hold for the naive estimator, that is, when $\hat F_n$ is replaced by $\tilde F_n$.
\end{theorem}

\begin{remark}
If $F_{0k}(s)>0$ for all $k=1,\dots,K$ and $\sum_{k=1}^K F_{0k}(s) =1$, then the matrix $V(s)$ is positive-semidefinite with rank $K-1$.
If $F_{0k}(s)=0$ or $F_{0k}(s)=1$, then the $k$th row and the $k$th column of $V(s)$ are zero vectors, and the corresponding limiting distributions of $\hat F_{nk}(s)$ and $\tilde F_{nk}(s)$ should be interpreted as degenerate distributions consisting of a point mass at zero. More details can be found in the proof of Theorem \ref{th: normal limit}.
\end{remark}

\subsection{Limiting distributions in the grouped model}\label{sec: large-sample grouped}

Denote the infimum and supremum of $\mathcal M$ by $m_{\inf}$ and $m_{\sup}$. Define 
$\{m(I)\}_- = \sup\{x\in \mathcal M \st x<m(I)\}$ for $I\in \mathcal I$ with $m(I)\neq m_{\inf}$, and
$\{m(I)\}_+ = \inf\{x\in \mathcal M \st x>m(I)\}$ for $I\in \mathcal I$ with $m(I)\neq m_{\sup}$.
If $\{m(I)\}_- \in \mathcal M$ let $I_- = I[\{m(I)\}_-]$, and if $\{m(I)\}_+ \in \mathcal M$ let  $I_+ = I[\{m(I)\}_+]$. We say that $I \in \mathcal I$ is a regular interval if $H_{0k}(I)=0$ for all $k=1,\dots,K$ or the following two conditions hold:
(i) if $m(I) \neq m_{\inf}$ then $\{m(I)\}_- \in \mathcal M$ and for each $k=1,\dots,K$ either $H_{0k}(I_-)<H_{0k}(I)$ or $H_{0k}(I)=0$, and (ii) if $m(I) \neq m_{\sup}$ then $\{m(I)\}_+ \in \mathcal M$ and for each $k=1,\dots,K$ either $H_{0k}(I) < H_{0k}(I_+)$ or $H_{0k}(I)=0$.

Analogously to $\breve F_n$ in the discrete model, we define a simple estimator in the grouped model:
   $$\breve H_{nk}(I) = M_{k}(I)/M(I), \qquad k=1,\dots,K, I\in \mathcal I.$$
The proofs and results for the discrete model can now be translated directly to the grouped model, by replacing regular points $s \in \mathcal S$ by regular intervals $I \in \mathcal I$, $\hat F_n(s)$ by $\hat H_n(I)$, $\tilde F_n(s)$ by $\tilde H_n(I)$, $\breve F_n(s)$ by $\breve H_n(I)$, $F_0(s)$ by $H_0(I)$, and $N_k(s)$ by $M_k(I)$ for $k=1,\dots,K+1$. We therefore only give the main result in Theorem \ref{th: normal limit grouped}, without proof.

\begin{theorem}\label{th: normal limit grouped}
   For any regular interval $I\in \mathcal I$ in the grouped model,
   \begin{align*}
      n^{1/2}\{ \hat {H}_n(I) - H_0(I)\} = n^{1/2} \left( \begin{array}{c}
          \hat {H}_{n1}(I)-H_{01}(I) \\
          \vdots\\
          \hat {H}_{nK}(I)-H_{0K}(I)
          \end{array}\right)
   \end{align*}
   is asymptotically normal with mean zero and covariance matrix $U(I)$, where
   $U(I)$ is a $K \times K$ matrix with entries
   \begin{align*}
      \{U(I)\}_{k,\ell} & = \left[ H_{0k}(I)\ind{k=\ell} - H_{0k}(I)H_{0\ell}(I) \right]/ G(I), \quad  k,\ell \in \{1,\dots,K\}.
   \end{align*}
  Moreover, for any finite collection of regular intervals $I_1,\dots,I_p$, the stacked vector
   $n^{1/2}\{ \hat {H}_n(I_1)-H_0(I_1), \dots, \hat {H}_n(I_p)-H_0(I_p)\}$
   is asymptotically normal with mean vector zero and block diagonal covariance matrix
   with blocks $U(I_1),\dots,U(I_p)$. Consistent estimators for the elements of $U(I)$, $I\in \mathcal I$, are
   \begin{align*}
      \{ \hat U_{n}(I)\}_{k,\ell} = \left[\hat {H}_{nk}(I)\ind{k=\ell} - \hat {H}_{nk}(I)\hat {H}_{n\ell}(I) \right]/M(I), \quad k,\ell\in \{1,\dots,K\}.
   \end{align*}
   The same results hold for the naive estimator, that is, when $\hat H_n$ is replaced by $\tilde H_n$.
\end{theorem}
As in Theorem \ref{th: normal limit}, a degenerate limiting distribution should be interpreted as point mass at zero.

\subsection{Theoretical motivation for the grouped model}

   The asymptotic results provide a theoretical motivation for the grouped model, since a contradiction arises with respect to rates of convergence when the grouping of observation times is ignored. To see this, consider the menopause data and the HIV data, and suppose that the grouping of observation times is ignored, meaning that the recorded observation times are interpreted as exact observation times. This assumption was made in previous analyses of the menopause data (see \cite{KrailoPike83, JewellVanderLaanHenneman03,
    JewellKalbfleisch04, Maathuis06thesis}).
     Under this assumption, the discrete model is most appropriate for the menopause data, since there are numerous ties in the recorded observation times, see Section \ref{sec: menopause}. On the other hand, the smooth model seems most appropriate for the HIV data, since this data set contains very few ties in the recorded observation times, see Section \ref{sec: HIV}. This would imply that the local rate of convergence of the maximum likelihood estimator and the naive estimator at the recorded observation times is $n^{1/2}$ for the menopause data, while it is $n^{1/3}$ for the HIV data.

    In reality, however, the observation times were continuous in both data sets, and they were rounded in the recording process. In the menopause data, this rounding was substantial, into 1-year or 5-year intervals, while in the HIV data it was minimal, into 1-day intervals. Since rounding implies discarding information, it seems impossible that more rounding, as in the menopause data, leads to a faster local rate of convergence at the recorded observation times.
    This apparent contradiction can be resolved by modeling the grouping of the observation times. For the grouped model, rounding or grouping of the observation times indeed yields a faster rate of convergence, but not for the cumulative incidence functions at the recorded observation times, but for weighted averages of the cumulative incidence functions over the grid cells. These weighted averages are smooth functionals of the cumulative incidence functions and thus can be estimated at rate $n^{1/2}$ (see \citet{JewellVanderLaanHenneman03}, \citet[Chapter 7]{Maathuis06thesis}).

\section{Construction of pointwise confidence intervals}
\label{sec: construction of confidence intervals}

\subsection{Confidence intervals in the discrete and grouped models}\label{sec: conf intervals discr grouped}
In the discrete and grouped models, the large-sample behavior of the maximum likelihood estimator and the naive estimator at regular points or intervals is standard,
and hence confidence intervals can be constructed
by any standard method, for example using the asymptotic normal
distribution or the bootstrap. For instance, let $s\in \mathcal S$ be a regular point in the discrete model. Then an asymptotic $(1-\alpha)100\%$ confidence interval for $F_{0k}(s)$
is \begin{align*}
   \hat F_{nk}(s) \pm n^{-1/2} z_{1-\alpha/2} [ \{ \hat V_n(s)\}_{k,k} ] ^{1/2},
\end{align*}
where $z_{1-\alpha/2}$ is the $(1-\alpha/2)$-quantile of the standard normal distribution.
Similarly, considering a regular interval $I\in \mathcal I$ in the grouped model, an asymptotic $(1-\alpha)100\%$ confidence interval for $H_{0k}(I)$ is
\begin{align}\label{eq: confidence interval grouped}
   \hat H_{nk}(I) \pm n^{-1/2} z_{1-\alpha/2} [ \{ \hat U_n(I)\}_{k,k} ]^{1/2}.
\end{align}

\subsection{Confidence intervals in the smooth model}\label{sec: conf intervals smooth}
In the smooth model, the large-sample behavior of the maximum likelihood estimator and the naive estimator is nonstandard, making the
construction of confidence intervals less straightforward.
In principle, one can construct confidence intervals using the limiting
distribution of the maximum likelihood estimator, but this approach entails several difficulties. First, the limiting distribution
involves parameters from the underlying distributions that need to be estimated. Moreover,
Theorems $1 \point 7$ and $1 \point 8$ of \citet{GroeneboomMaathuisWellner08b} suggest that
these parameters cannot be separated from the limiting
distribution, in the sense that it seems impossible to write the limiting distribution as $c Z$,
where $c$ is a constant depending on the underlying distribution and $Z$ is a universal limit.
Hence, one would need to simulate the limiting
distribution on a case by case basis. Conducting such simulations is non-trivial \citep{GroeneboomWellner01}.

One might also consider the nonparametric bootstrap to construct confidence intervals based on the maximum likelihood estimator or the naive estimator. However, it is likely the bootstrap is inconsistent in this setting, given recent results of \citet{Kosorok08} and \citet{SenBanerjeeWoodroofe10} on inconsistency of the bootstrap for the closely related Grenander estimator.

Subsampling \citep{PolitisRomano94}, a variant of the bootstrap, produces asymptotically valid confidence intervals under minimal assumptions, and can be applied to construct asymptotically valid confidence intervals for the cumulative incidence functions based on the maximum likelihood estimator or the naive estimator. A drawback of subsampling is that it requires a tuning parameter, the subsample size, which is difficult to choose in practice.

Finally, one can consider likelihood ratio confidence intervals based on the naive estimator. Although the naive estimator has been shown empirically to be less efficient than the maximum likelihood estimator \citep[Figure 3]{GroeneboomMaathuisWellner08b}, it has the advantage that its large sample behavior is simpler. For a fixed failure cause, the limiting distribution of the naive estimator is identical to the limiting distribution of the maximum likelihood estimator for current status data without competing risks \citep[Theorem $1\point 6$]{GroeneboomMaathuisWellner08b}. Hence, the likelihood ratio theory of \citet{BanerjeeWellner01} applies, and confidence intervals can be constructed by inverting likelihood ratio tests \citep{BanerjeeWellner05}. These confidence intervals have the appealing property that they do not require estimation of parameters from the underlying distribution, nor any tuning parameters. Simulation studies by \citet{BanerjeeWellner05} showed that for current status data without competing risks, likelihood ratio based confidence intervals are typically preferable over confidence intervals based on the limiting distribution or subsampling. In the smooth model, we therefore recommend using likelihood ratio confidence intervals based on the naive estimator.

\section{Examples}
\label{sec: examples}

\subsection{Simulation}\label{sec: simulation}

It is not clear how well the asymptotic distributions of Sections \ref{sec: large-sample local discrete} and \ref{sec: large-sample grouped} approximate the finite sample behavior of the estimators, especially for grids that are dense relative to $n$. We therefore conducted a simulation study, using the following discrete model: $\P(Y=1)=0 \point 6,$ $\P(Y=2)=0 \point 4$, $X \mid Y=1 \sim \text{Gamma}(5,3)$, and $X \mid Y=2 \sim \text{Gamma}(9,2)$. The distribution of $C$ was uniform on one of the following grids: (i) $\{10,20,30\}$, called gap $10$, (ii) $\{6,8,\dots,34\}$, called gap $2$, (iii) $\{5\point 5,6 \point 0,\dots,35 \point 0\}$, called gap $0 \point 5$, and (iv) $\{5 \point 1,5 \point 2,\dots,35 \point 0\}$, called gap $0 \point 1$. For each of the four resulting models, 1000 data sets of sample size $n=1000$ were simulated. Symmetric $95\%$ asymptotic confidence intervals for the cumulative incidence functions were computed at the points $t_0 = (10,20, 30)$, based on the normal distribution and the bootstrap, using both the maximum likelihood estimator and the naive estimator.

\begin{figure}[!htp]
     \vspace{-.4cm}
     \hspace{-1.8cm}
        \includegraphics[scale=.78,angle=-90]{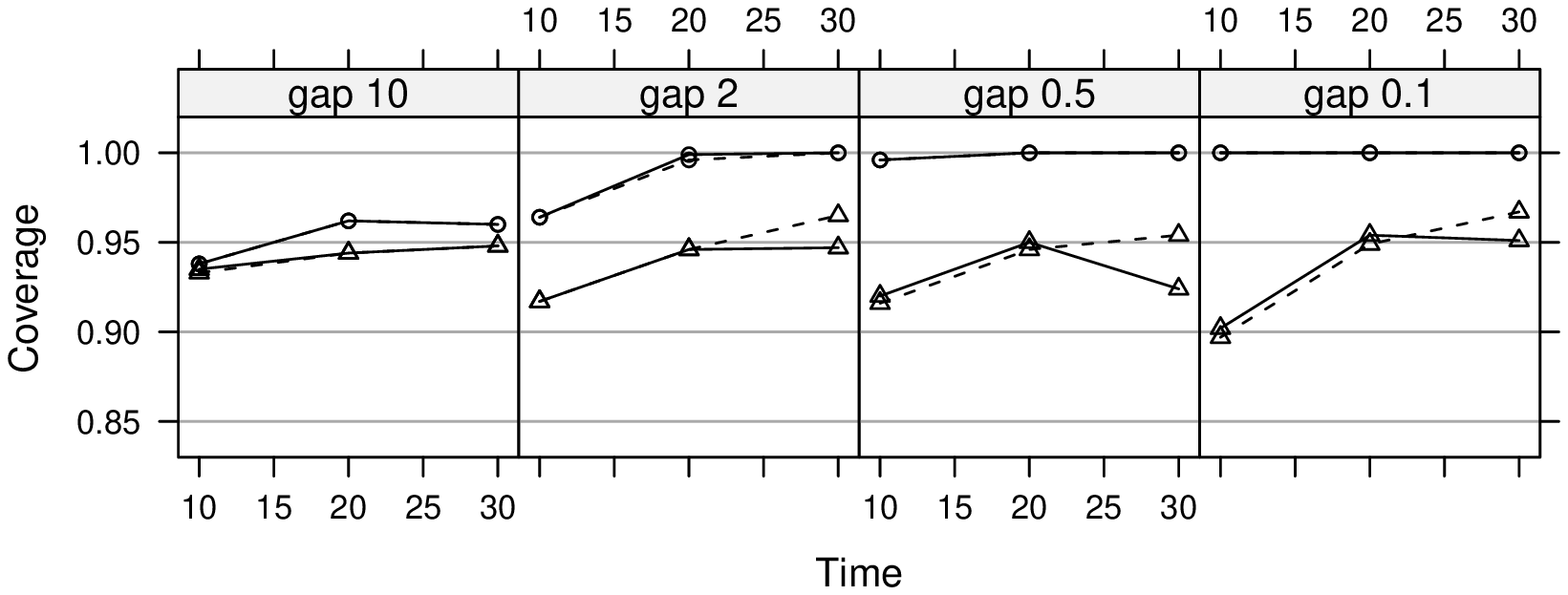}\\[-1cm]
     \hspace*{-1.8cm}
        \includegraphics[scale=.78,angle=-90]{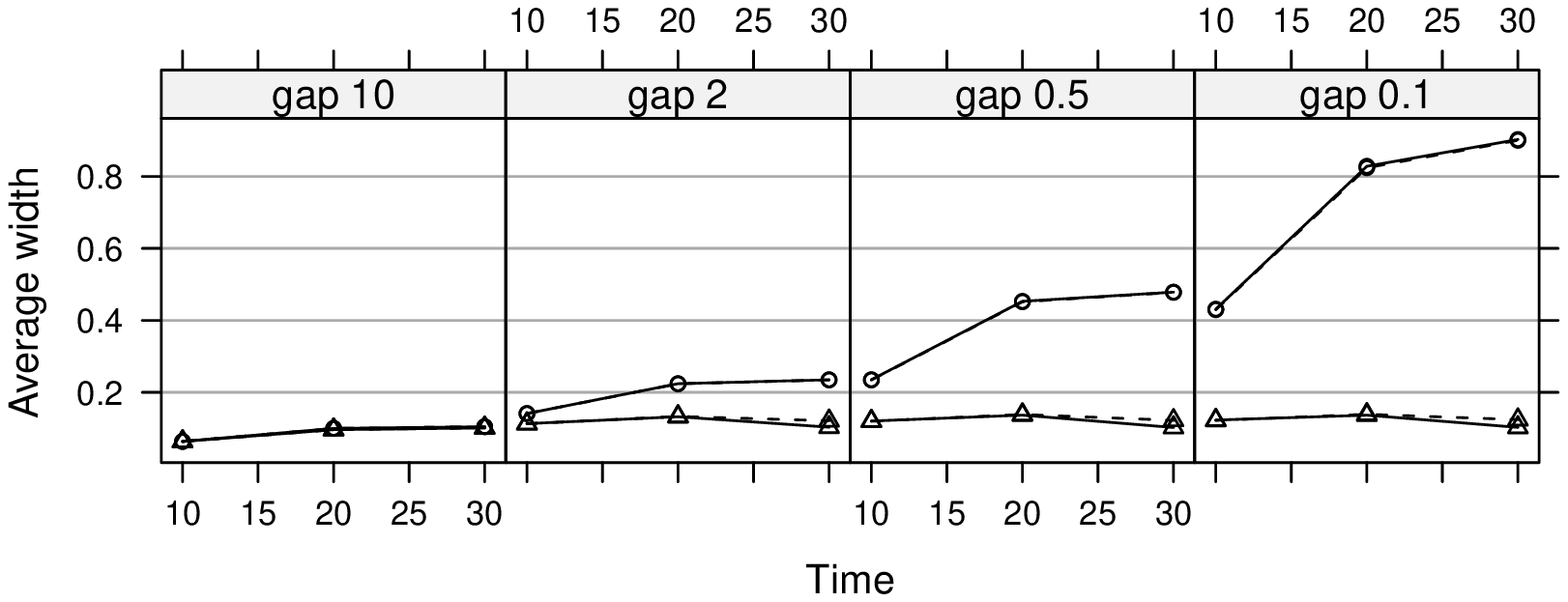}
     \vspace{-.7cm}
  \caption{Simulation: Coverage and average width of the four $95\%$ confidence intervals for $F_{01}(t_0)$ as a function of $t_0$. The confidence intervals were based on the normal distribution $(\circ)$ and the bootstrap $(\vartriangle)$, using the maximum likelihood estimator (solid line) and the naive estimator (dashed line). The bootstrap confidence intervals are based on 750 bootstrap samples.}
     \label{fig: simul}
\end{figure}

The results for $F_{01}$ are shown in Figure \ref{fig: simul}. The results for $F_{02}$ are similar, and are therefore omitted. Confidence intervals based on the maximum likelihood estimator behave very similarly to confidence intervals based on the naive estimator, while there is a large difference between normal and bootstrap based confidence intervals for the denser grids. The increase in width of the normal based confidence intervals for the denser grids is caused by the decrease of $nG(\{t_0\})$, which can be viewed as the expected effective sample size for the simple estimator $\breve F_n$ at $t_0$. As a result, the variance of the asymptotic normal distribution increases by a factor 5 or 6 between each pair of successive grids. The empirical variance of the estimators, on the other hand, increases somewhat for the denser grids, but not by much, due to the stabilizing effect of the monotonization that takes place in the maximum likelihood estimator and the naive estimator. As a result, the normal based confidence intervals give substantial over-coverage. This breakdown of the normal limit is already apparent for the larger time points in the relatively coarse grid gap 2, which has an average of 67 observation times per grid point. The bootstrap variance was found to be a better approximation of the empirical variance of the estimators, suggesting the use of bootstrap intervals over asymptotic normal intervals in practice. However, the under-coverage of the bootstrap intervals at $t_0=10$ becomes more substantial as the grids become denser. This points to inconsistency of the bootstrap for very dense grids, which is in line with the theory discussed in Section \ref{sec: conf intervals smooth}.

\subsection{Menopause data}\label{sec: menopause}

   We consider data on 2423 women in the age range 25-59 years from Cycle I
   of the Health Examination Survey of the National Center for Health
   Statistics \citep{MacMahonWorcestor66}. Among other things, these women were asked to report their current age and whether or not they were post-menopausal. Moreover, if they were postmenopausal, they were asked to report the age and cause of menopause, where the cause could be natural or operative. Since \citet{MacMahonWorcestor66} found marked terminal digit clustering in the reported ages of
   menopause, \citet{KrailoPike83} excluded these
   from the analysis. The remaining information can be viewed as
   current status data with competing risks.
   Nonparametric estimates of the cumulative incidences of the two types of menopause
   were computed by
   \citet{JewellVanderLaanHenneman03}, \citet{JewellKalbfleisch04} and
   \citet{Maathuis06thesis} under the assumption that the recorded ages of the women at the time of the interview were exact. However, this was not the case. Instead, the ages were grouped into the intervals $(25,30]$, $(30,35]$, $(35,36]$, $(36,37] \dots,(58,59]$ and recorded as the midpoints of these intervals, yielding 26 age groups with a minimum of 45 and an average of 93 observations per age group. This is comparable to gap 2 in our simulation study, see Section \ref{sec: simulation}.

\begin{figure}[!htp]
     \centering
     \subfigure{
        \includegraphics[scale=.46,angle=-90]{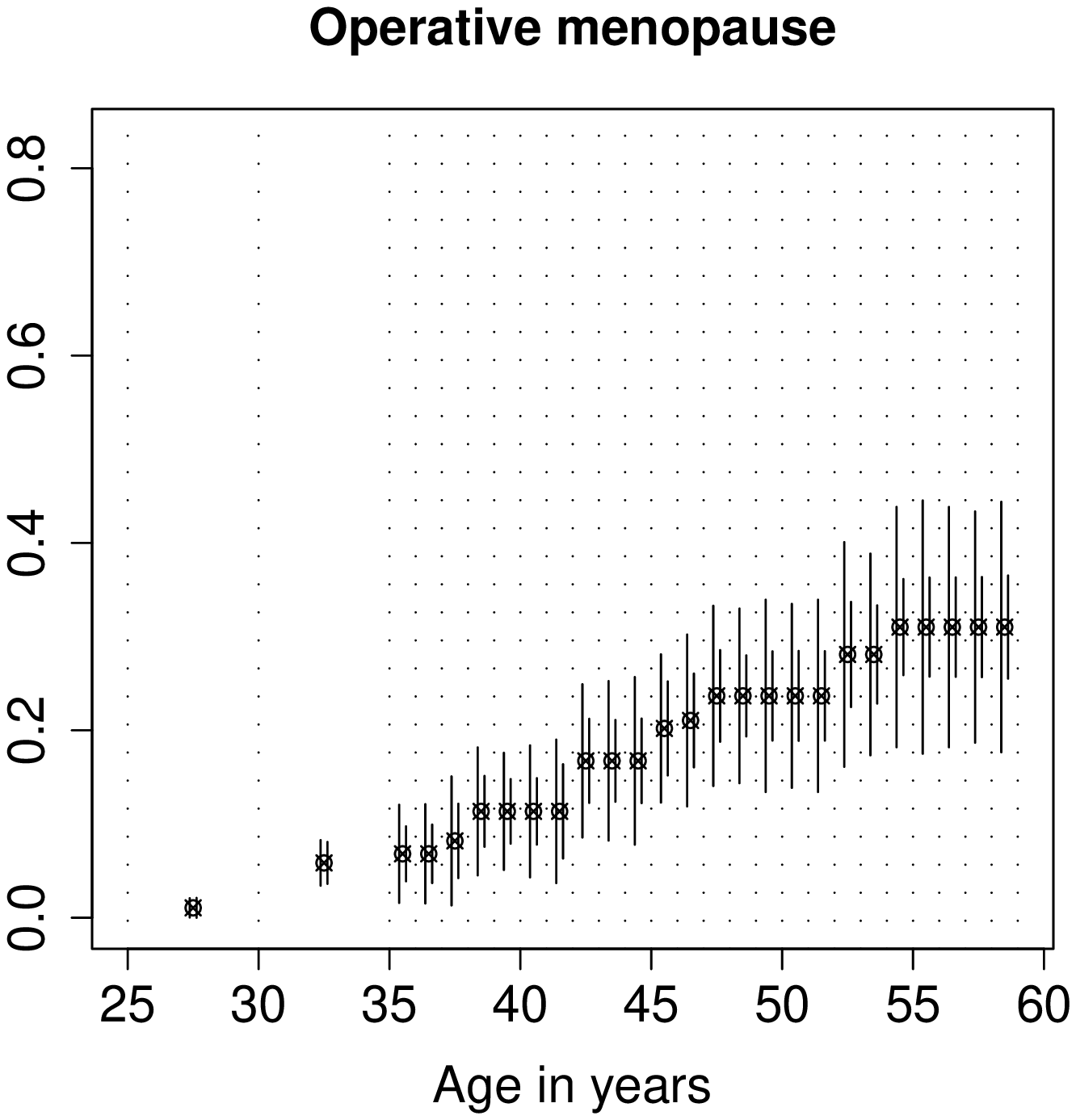}
     }
     \subfigure{
        \includegraphics[scale=.46,angle=-90]{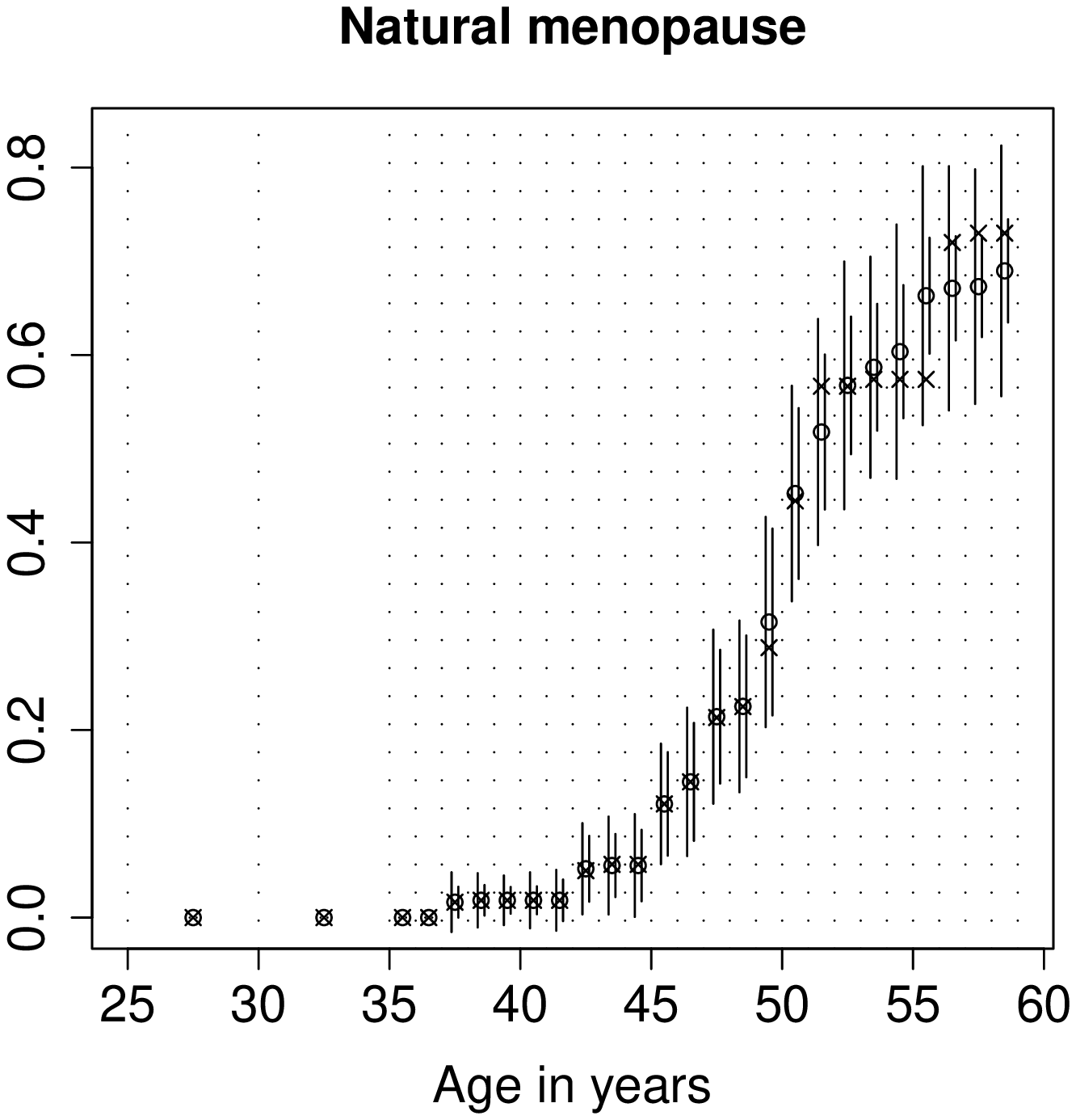}
     }
  \vspace{-.7cm}
  \caption{Menopause data: The maximum likelihood estimator $\hat H_n$ $(\circ)$ and the naive estimator $\tilde H_n$ $(\times)$ for the weighted averages of the cumulative incidence of operative and natural menopause over the age groups. The estimators are plotted at the midpoints of the age groups which are indicated by the dotted vertical lines. The two solid vertical line segments in each age group are $95\%$ asymptotic confidence intervals based on the maximum likelihood estimator: the left line segment is based on the normal approximation \eqref{eq: confidence interval grouped} and the right line segment is a symmetric bootstrap confidence interval based on 1000 bootstrap samples.}
  \label{fig: menopause}
  \vspace{-.8cm}
\end{figure}

   We add to the previous analyses of these data in two ways: we use the grouped model, which is clearly appropriate for these data, and we provide confidence intervals. Figure \ref{fig: menopause} shows the maximum likelihood estimator and the naive estimator
    for the weighted averages of the cumulative incidence functions, together with $95\%$ normal and bootstrap confidence intervals based on the maximum likelihood estimator. As in our simulation study, the confidence based on the normal distribution are wider than those based on the bootstrap.

\subsection{HIV data}\label{sec: HIV}

\begin{figure}
  \centering
  \subfigure{
     \includegraphics[scale=.45,angle=-90]{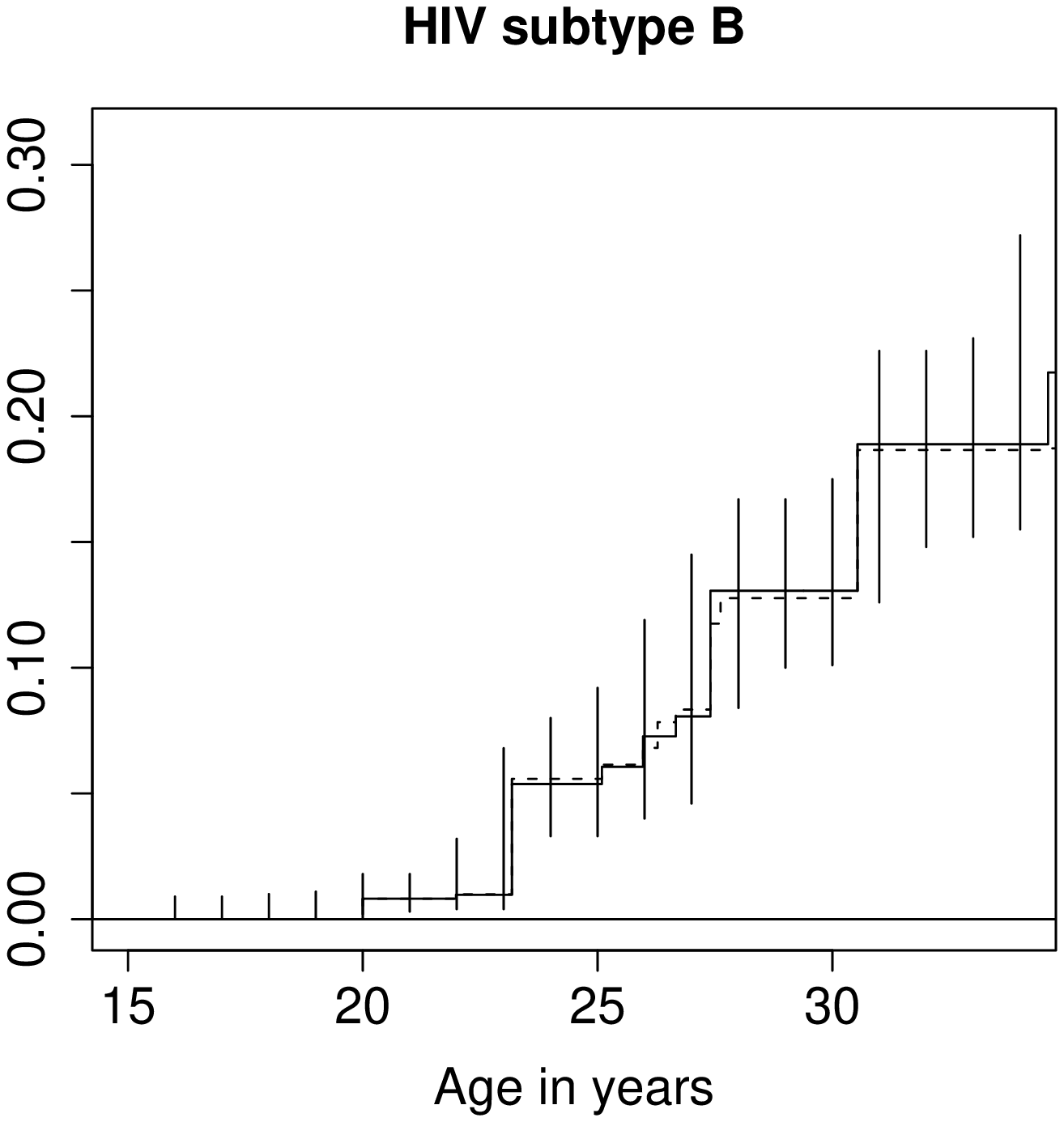}
  }
  \subfigure{
      \includegraphics[scale=.45,angle=-90]{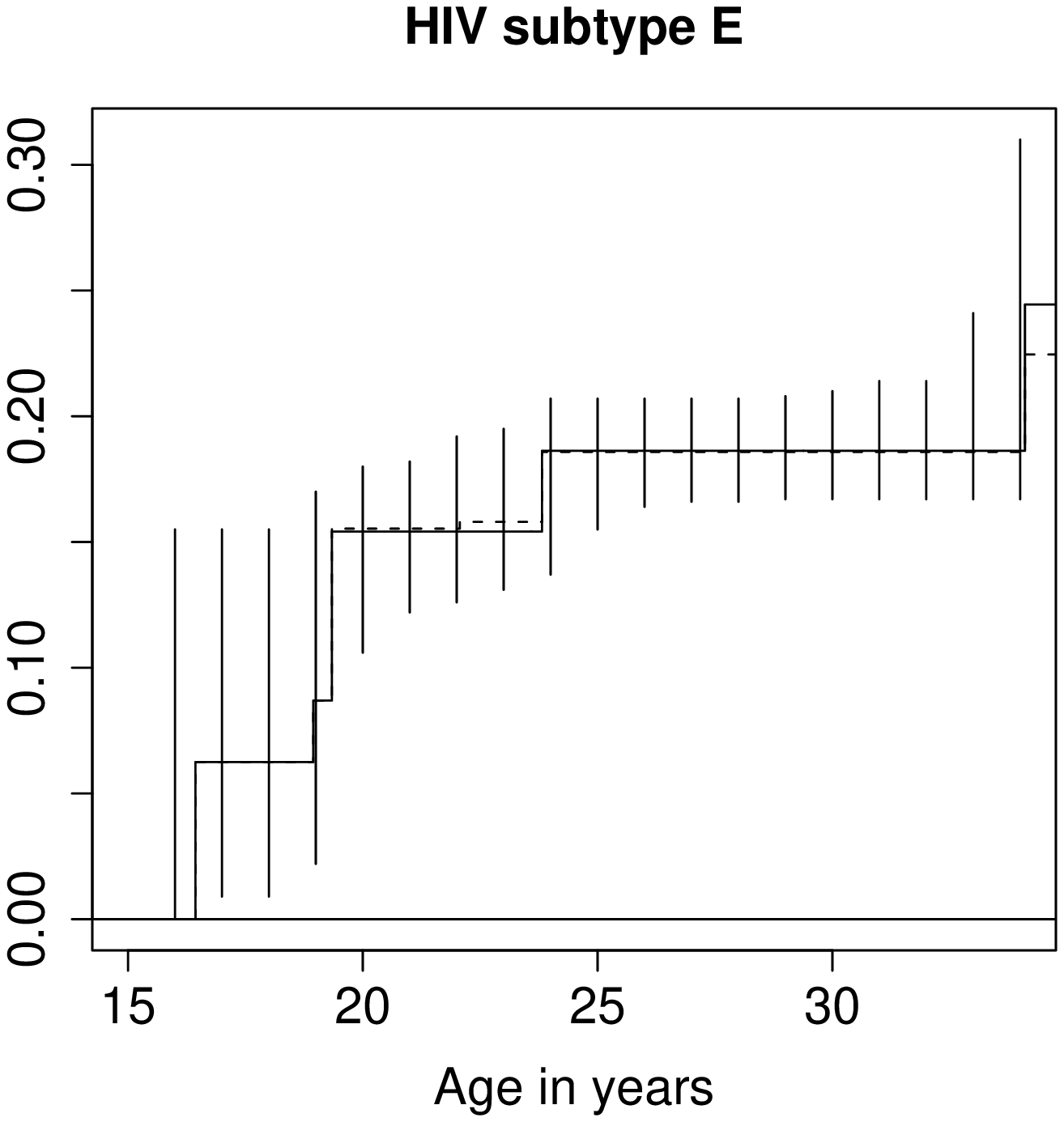}
  }
  \vspace{-.7cm}
  \caption{HIV data: The maximum likelihood estimator $\hat F_n$ (dashed) and the naive estimator $\tilde F_n$ (solid) for the cumulative incidence of HIV subtypes B and E as a function of age, using the smooth model. The solid vertical lines represent $95\%$ pointwise confidence intervals at times $16,\dots,34$, based on the likelihood ratio method for the naive estimator.}
  \label{fig: HIV}
\end{figure}

The Bangkok Metropolitan Administration injecting drug users cohort study \citep{KitayapornEtAl98, VanechseniEtAl01} was established in 1995 to
better understand HIV transmission and to assess the feasibility of conducting a phase III HIV
vaccine efficacy trial in an injecting drug users population in Bangkok.
We consider data on $1366$ injecting drug users in this study who were screened from May to December 1996 and who were under 35 years of age.
Among this group, 393 were HIV positive, with
114 infected with subtype B, 238 infected with subtype E, 5 infected by another or mixed subtype, and 36
infected with missing subtype. The subjects with other, mixed, or missing subtypes were grouped in a remainder category. All ages were recorded in days, leading to a small number of ties:
among the 1366 subjects, there were 1212 distinct ages, and the mean number of observations per distinct
age was 1.13. In light of this, we analyze these data using the smooth model. Figure \ref{fig: HIV} shows the maximum likelihood estimator and the naive estimator for the subtype-specific cumulative incidence of HIV, together with $95\%$ likelihood ratio confidence intervals based on the naive estimator.

\section{Observation time distribution or grouping dependent on $n$}

There are interesting connections between our work and unpublished work of Tang, Banerjee and Kosorok (see http://www.stat.lsa.umich.edu/$\sim$moulib/jsm09csd.pdf), who studied current status data without competing risks when the observation time distribution depends on the sample size $n$. More precisely, let $X$ be a random event time with distribution $F_0$ and let $C^{(n)}$ be a random observation time with distribution $G^{(n)}$, where $G^{(n)}$ is a discrete distribution on an equidistant grid with spacings $n^{-\gamma}$ for some  $\gamma\in (0,1)$. Without loss of generality, assume this grid is on $[0,1]$. Consider the nonparametric maximum likelihood estimator $\hat F_n$ for $F_0$ based on $n$ independent and identically distributed observations of $(C^{(n)},\Lambda^{(n)})$, where $\Lambda^{(n)} = 1\{X\le C^{(n)}\}$. Let $t_0\in (0,1)$ be a time point of interest, and let $t_{n}$ be the largest support point of $G^{(n)}$ smaller than $t_0$.
Assuming $F_0$ satisfies certain smoothness conditions in a neighborhood of $t_0$, Tang \emph{et al.} found that the limiting distribution of the maximum likelihood estimator depends crucially on $\gamma$. For $\gamma<1/3$ the limiting distribution of $n^{(1-\gamma)/2} \{ \hat F_n(t_{n})-F_0(t_{n})\}$ is normal with mean zero and variance $F_0(t_0)\{1-F_0(t_0)\}$. Hence, for such sparse grids, the maximum likelihood estimator behaves as in the discrete model, up to a different rate of convergence. For  $\gamma>1/3$, on the other hand, the limiting distribution of $n^{1/3}\{\hat F_n(t_0)-F_0(t_0)\}$ is determined by the slope of the convex minorant of a Brownian motion process plus parabolic drift, showing that the maximum likelihood estimator behaves as in the smooth model. The case $\gamma=1/3$ forms the boundary between these two scenarios and yields a new limiting distribution.

Combining our work with that of Tang \emph{et al.} yields two extensions. First, consider a grouped model for current status data without competing risks, where the grouping intervals depend on $n$. More precisely, let $X$ be an event time with distribution $F_0$, let $C$ be an observation time with distribution $G$, and let $\Lambda = 1\{X\le C\}$. Assume the support of $G$ is $[0,1]$, and let $\mathcal I_n$ be the set of intervals formed by  the grid cells of an equidistant grid on $[0,1]$ with spacings $n^{-\gamma}$ for some $\gamma \in (0,1)$. Assume that the observation time $C$ is rounded to the midpoint of the interval in which it falls, and denote this rounded observation time by $D^{(n)}$. One can now consider the nonparametric maximum likelihood estimator for $F_0$ based on $n$ independent and identically distributed copies of $(D^{(n)},\Lambda)$. Since the likelihood in this grouped model can be written in exactly the same form as the likelihood in the discrete model, and since also the constraints on the two optimization problems are equivalent, the work of Tang \emph{et al.} should carry over to this model, with the only difference that everything should written in terms of weighted averages of $F_0$ over the grid cells.
Second, consider the discrete model for current status data with competing risks, where the support of $G$ depends on $n$. Then the results of Tang \emph{et al.} should carry over to the naive estimator $\tilde F_{nk}$, since this estimator can viewed as a maximum likelihood estimator based on reduced current status data without competing risks. The same holds for the naive estimator $\tilde H_{nk}$ in the grouped model when the grouping intervals depend on $n$.


\section*{Acknowledgements}

We thank Kachit Choopanya, Dwip Kitayaporn, Timothy D.
Mastro, Philip A. Mock and Suphak Vanichseni for allowing us to use the
data from the Bangkok Metropolitan Administration Injecting Drug Users cohort study. We thank Moulinath Banerjee for discussing the connections between this paper and his work with Runlong Tang and Michael Kosorok. We thank the Associate Editor and two anonymous referees for helpful suggestions and comments. Michael G. Hudgens was supported by a grant from the National Institutes of Health.

\appendix
\section*{Appendix 1}
\subsection*{Proofs}\label{sec: proofs}

\begin{proof_of}[Lemma \ref{lemma: breve Fn}]
   Due to the absence of monotonicity constraints on $\mathcal F_K^*$, the maximizer of $l_n(F)$ over $\mathcal F_K^*$ can be determined separately for each $s \in \mathcal S$. Thus, fix $s\in \mathcal S$, and define $l_n(F,s) = \sum_{k=1}^{K+1} N_k(s)\log \{F_k(s)\}$. Moreover, define $\mathcal K = \{k\in \{1,\dots,K+1\} \st N_k(s)>0\}$ and $\mathcal K^C = \{1,\dots,K+1\}\setminus \mathcal K$.
   First, suppose $\mathcal K = \emptyset$. Then $l_n(F,s)=0$ for any choice of $F_k(s)$, $k=1,\dots,K$, and hence $\breve F_n(s)$ is a maximizer of $l_n(F,s)$.
   Next, suppose $\mathcal K \neq \emptyset$, or equivalently, $N(s)>0$. Then any maximizer of $l_n(F,s)$ subject to the constraint $F_+(s) \le 1$ must set $F_k(s)=0$ for $k\in \mathcal K^C$. Hence, for $k\in \mathcal K^C$ the maximizer is unique and equals $\breve F_{nk}(s)$. If $|\mathcal K|=1$, $l_n(F,s)$ contains only one non-zero term, and it is clear that the corresponding $F_k(s)$ should be set to 1, which equals $\breve F_{nk}(s)$. If $|\mathcal K|>1$, we define $k^* = \max \mathcal K$. Then $N_{k^*}(s) = N(s) - \sum_{k\in \mathcal K \setminus \{k^*\}}N_k(s)$ and any maximizer of $l_n(F,s)$ over $\mathcal F_K^*$ must satisfy $F_{k^*}(s) = 1-\sum_{k\in \mathcal K \setminus\{k^*\}} F_k(s)$. Hence, we can write $l_n(F,s) = \sum_{k\in \mathcal K \setminus \{k^*\}} N_k(s)\log\{F_k(s)\} + \{N(s)-\sum_{k\in \mathcal K \setminus \{ k^*\}}N_k(s)\}\log\{1-\sum_{k\in \mathcal K \setminus \{k^*\}}F_k(s)\}$. This function is strictly concave in $F_k(s)$ for $k\in \mathcal K \setminus \{k^*\}$. The unique maximizer can be determined by solving $\partial l_n(F,s) / \partial F_k(s) = 0$ for $k \in \mathcal K \setminus \{k^*\}$, which yields $\breve F_{nk}(s)$, $k\in \mathcal K$. \qed
\end{proof_of}

     \begin{proof_of}[Lemma \ref{th: MLE=multinomial}]
         Let $s\in \mathcal S$ be a regular point in the discrete model. We first consider the maximum likelihood estimator for the basic case where $s\notin\{s_{\inf}, s_{\sup}\}$. Let $\mathcal K^+ = \{ k \in \{1,\dots,K\} \st F_{0k}(s)>0\}$.
         For $k\in \{1,\dots,K\}\setminus \mathcal K^+$, we have $N_k(s)=0$. Hence, the corresponding $F_k(s)$'s do not contribute to the likelihood and we directly obtain that the corresponding estimators satisfy $\hat F_{nk}(s)=\breve F_{nk}(s)=0$. So we are done if $\mathcal K^+ = \emptyset$. Otherwise, we are left to show $\P[ \cap_{k\in \mathcal K^+} \{ \hat F_{nk}(s)=\breve F_{nk}(s)\}] \to 1$ as $n\to \infty$.
         Define the events
         \begin{align*}
            A_n(s) & = \cap_{k \in \mathcal K^+} \{ \hat F_{nk}(s_-) < \hat F_{nk}(s) < \hat F_{nk}(s_+)\},\\
            B_n(s) & = \cap_{k \in \mathcal K^+ \cup \{K+1\}} \{N_k(s)>0\}.
         \end{align*}
         The assumptions on $s$ imply $F_{0k}(s_-)<F_{0k}(s)<F_{0k}(s_+)$ for $k \in \mathcal K^+$. By combining this with the consistency of $\hat F_n$ (Theorem \ref{th: consistency discrete}), it follows that $\P\{ A_n(s) \} \to 1$ as $n\to \infty$. Moreover, the law of large numbers, $G(\{s\})>0$, $F_{0k}(s)>0$ for $k\in \mathcal K^+$, and $F_{0+}(s)<1$ imply $\P\{ B_n(s) \} \to 1$ as $n\to \infty$. Hence, $\P\{A_n(s)\cap B_n(s)\} \to 1$ as $n \to \infty$, and the proof for the basic case can be completed by showing that the event $\{A_n(s) \cap B_n(s)\}$ implies  $\cap_{k\in \mathcal K^+} \{ \hat F_{nk}(s)=\breve F_{nk}(s)\}$. We do this using contraposition. Thus, suppose $\{A_n(s)\cap B_n(s)\}$ holds. This implies $k^*=K+1$ in the proof of Lemma \ref{lemma: breve Fn}, and it follows that $\breve F_{nk}(s)$, $k\in \mathcal K^+$, is the unique solution of $\partial l_n(F)/\partial F_k(s) = 0$, $k\in \mathcal K^+$. Now assume there is a $j\in \mathcal K^+$ such that $\hat F_{nj}(s) \neq \breve F_{nj}(s)$. Then there must be a $\bar k \in \mathcal K^+$ such that $\partial l_n(F)/\partial F_{\bar k}(s)|_{\hat F_n(s)} \neq 0$. Let $\sigma \in \{-1,+1\}$ be the sign of $\partial l_n(F)/\partial F_{\bar k}(s)|_{\hat F_n(s)}$, and define $\hat F_n^{new}(s) = \hat F_n(s) + \gamma \sigma e_{\bar k}$, where $e_{k}$ is the unit vector in $\R^K$ with a $1$ at the $k$th entry. Then for $\gamma>0$ sufficiently small, replacing $\hat F_n(s)$ by $\hat F_{n}^{new}(s)$ increases the log likelihood. Moreover, this replacement does not violate the constraints of $\mathcal F_K$, as for $\gamma>0$ sufficiently small we have $\hat F_{n\bar k}(s_-) < \hat F_{n\bar k}^{new}(s) < \hat F_{n \bar k}(s_+)$ and $\hat F_{n+}^{new}(s) < \hat F_{n+}(s_+) \le 1$. This shows that $\hat F_{n}$ cannot be the maximum likelihood estimator, which is a contradiction.

         If $s\neq s_{\inf}$ and $s=s_{\sup}$, we distinguish two cases. If $F_{0+}(s)<1$, the proof of the basic case goes through with the only change that $A_n(s) = \cap_{k\in \mathcal K^+} \{\hat F_{nk}(s_-) < \hat F_{nk}(s)\} \cap \{\hat F_{n+}(s)<1\}$. If $F_{0+}(s)=1$, then $N_{K+1}(s)=0$ and $1-F_+(s)$ does not contribute to the log likelihood. Hence, the maximum likelihood estimator must satisfy $\hat F_{n+}(s)=1$ and this equals $\breve F_{n+}(s)$ if $N(s)>0$. If $K=1$, this implies $\P\{\hat F_n(s) = \breve F_n(s)\} \to 1$ as $n\to \infty$, so that we are done. If $K>1$, we use the proof for the basic case with the following changes. We define $A_n(s) = \cap_{k\in \mathcal K^+} \{\hat F_{nk}(s_-) < \hat F_{nk}(s)\}$ and $B_n(s) = \cap_{k\in \mathcal K^+} \{N_k(s)>0 \}$. As before, we have $\P\{A_n(s) \cap B_n(s)\} \to 1$ as $n\to \infty$. We will therefore show that $\{A_n(s)\cap B_n(s)\}$ implies $\cap_{k\in \mathcal K^+} \{\hat F_{nk}(s)=\breve F_{nk}(s)\}$, using contraposition. Thus, assume $\{A_n(s)\cap B_n(s)\}$ holds. This implies $k^* = \max \mathcal K^+$ in the proof of Lemma \ref{lemma: breve Fn}, meaning that $\breve F_{nk}$, $k\in \mathcal K^+$, are found by solving $\partial l_n(F) / \partial F_k(s)=0$ for $k\in \mathcal K^+ \setminus \{k^*\}$ and setting $\breve F_{nk^*}(s) = 1-\sum_{k\in \mathcal K^+ \setminus \{k^*\}} \breve F_{nk}(s)$. Assume $\hat F_{nk}(s) \neq \breve F_{nk}(s)$ for some $k\in \mathcal K^+$. Then there must be a $\bar k \in \mathcal K^+ \setminus \{k^*\}$ such $\partial l_n(F) / \partial F_{\bar k}(s)|_{\hat F_n(s)} \neq 0$. Define $\sigma$ as the sign of $\partial l_n(F) / \partial F_{\bar k}(s)|_{\hat F_n(s)}$, and define $\hat F_n^{new}(s) = \hat F_n(s) + \gamma \sigma e_{\bar k}  - \gamma \sigma e_{k^*}$. Then for $\gamma>0$ sufficiently small, replacing $\hat F_n(s)$ by $\hat F_{n}^{new}(s)$ increases the log likelihood. Moreover, this replacement does not violate the constraints of $\mathcal F_K$, as for $\gamma>0$ sufficiently small we have $\hat F_{n\bar k}(s_-) < \hat F_{n\bar k}^{new}(s)$, $\hat F_{n k^*}(s_-) < \hat F_{n k^*}^{new}(s)$, and $\hat F_{n+}^{new}(s) = \hat F_{n+}(s) =1$.
         Hence, $\hat F_n$ cannot be the maximum likelihood estimator, and we have again derived a contradiction.

         The proof for the maximum likelihood estimator is completed by considering two remaining special cases.
         If $s=s_{\inf}$ and $s\neq s_{\sup}$, then the proof for the basic case goes through with the only change that $A_n(s) = \cap_{k\in \mathcal K^+} \{0 < \hat F_{nk}(s) < \hat F_{nk}(s_+)\}$. If $s=s_{\inf}=s_{\sup}$, then $|\mathcal S|=1$ and monotonicity constraints do not play any role in the maximum likelihood estimator, so that $\hat F_n = \breve F_n$ follows immediately.

         The proof for the naive estimator follows directly from the proof for the maximum likelihood estimator by taking $K=1$. To see this, let $k\in \{1,\dots,K\}$ and recall that the naive estimator is the maximum likelihood estimator for the reduced current status data $(\Delta_k^i, C_i)$, $i=1,\dots,n$. Hence, the proof for the maximum likelihood estimator implies $\P\{\tilde F_{nk}(s) = \breve F_{nk}^{red}(s)\} \to 1$ as $n\to \infty$, where $\breve F_{nk}^{red}$ is the simple estimator based on the reduced data. The proof is completed by observing that $\breve F_{nk}^{red} = \breve F_{nk}$.
     \end{proof_of}

     \begin{proof_of}[Theorem \ref{th: normal limit}]
        Because of Lemma \ref{th: MLE=multinomial}, it is sufficient to derive the limiting distribution of $\breve F_n$. Let $k\in \{1,\dots,K\}$ and $s\in \mathcal S$. Since $\P\{N(s)>0\} \to 1$ as $n\to \infty$, we can assume $N(s)>0$. We first consider the case $0<F_{0k}(s)<1$. Then
        \begin{align*}
           n^{1/2} \{\breve F_{nk}(s) - F_{0k}(s)\} & = N(s)^{-1} n^{1/2} \{N_{k}(s)-F_{0k}(s)N(s)\} \\
                                                       & = N(s)^{-1} n^{-1/2} \sum_{i=1}^n \{\Delta_k^i - F_{0k}(s)\}\ind{C_i=s},
        \end{align*}
         and the result follows from $N(s) \convprob G(\{s\})$, the multivariate central limit theorem, and Slutsky's lemma (e.g., \citet[Lemma $2 \point 8$ (iii)]{VanderVaart98}).

         If $F_{0k}(s)=0$, then $N_k(s)=0$ and hence $\breve F_{nk}(s)=0=F_{0k}(s)$ always. Similarly, if $F_{0k}(s)=1$, we have $N_k(s)=N(s)$ and hence $\breve F_{nk}(s)=1=F_{0k}(s)$ whenever $N(s)>0$. These results are in agreement with the theorem, since in these cases $\{V(s)\}_{k,k}=0$, leading to a degenerate limiting distribution that should be interpreted as a point mass at zero. It can be easily verified that the off-diagonal elements $\{V(s)\}_{k,j}=0$ for $j\in \{1,\dots,K\}$, $j\neq k$, are also correct in these cases.
         \qed
     \end{proof_of}

\bibliographystyle{biometrika}
\bibliography{Mybibliography}

\end{document}